\documentstyle[12pt,epsfig]{article}
\begin{document}
\oddsidemargin 0mm
\evensidemargin 0mm
\voffset-20mm
\hoffset10mm
\pagestyle{plain}
\begin{center}\Large\bf
%-----------------------------------------------------------------
%  Title
%-----------------------------------------------------------------
X-ray properties of three clusters of galaxies and their implications for the Sunyaev-Zeldovich effect
%-----------------------------------------------------------------
\end{center}
\begin{center}
%-----------------------------------------------------------------
%  Authors
%-----------------------------------------------------------------
Michael Thierbach, Reinhard Schlickeiser, and Richard Wielebinski
%-----------------------------------------------------------------
\end{center}
\begin{center}
%-----------------------------------------------------------------
%  Institutes
%-----------------------------------------------------------------
Max-Planck-Institut f{\"u}r Radioastronomie, Postfach 2024, 53010 Bonn, Germany
%-----------------------------------------------------------------
\end{center}
%-----------------------------------------------------------------
%  Abstract here: maximum 200 words
%-----------------------------------------------------------------

\begin{abstract}
We present the preparation for an observation of the Sunyaev-Zeldovich (SZ) effect with the 100m telescope in Effelsberg. We calculate the expected diminution of the cosmic microwave background radiation (CMB) at radio wavelength towards a small sample of clusters of galaxies: A85, A665, and Cl0016+16. We obtain the required parameters from the X-ray analysis of ROSAT PSPC data of these clusters. We derive expected central diminutions $\Delta T_{{\rm R,max}}$ of $-(440\pm 180)\,\mu$K for A665 and $-(810\pm 680)\,\mu$K for Cl0016+16. For the cluster A85 we find evidence for a cooling flow. The best candidate for a SZ measurement is the cluster 0016+16.
\end{abstract}

%-----------------------------------------------------------------
%  text
%-----------------------------------------------------------------

%-----------------------------------------------------------------
%  end of abstract
%-----------------------------------------------------------------

\section{Introduction}

Sunyaev and Zeldovich (1972, 1980) showed that the passage of the photons of the CMB through clusters of galaxies distorts the spectrum of this radiation via the inverse Compton interaction. This distortion can be described by a change of the brightness temperature $T_{\rm R}$ of the CMB. This is known as the so-called Sunyaev-Zeldovich effect. It can be used as an important cosmological probe (e.g.~determination of the Hubble constant $H_{0}$). For a recent review of the SZ effect see Rephaeli (1995).

The strength of the SZ effect is determined by the properties of the intracluster electron distribution. This distribution is also the source of the X-ray emission of clusters of galaxies. With an analysis of this X-rays we are able to obtain the required electron parameters. To describe the distribution of the intracluster electrons we assume the well-known isothermal $\beta$-model, where the temperature of the electrons is taken as constant over the whole cluster and the density distribution is described by a King profile. Moreover, we assume only nonrelativistic electron temperatures.

\section{ROSAT observations}

We analyze the X-ray emission of the two Abell clusters A85 and A665, and of the cluster 0016+16. The relevant observations were made in 1991 and 1992 by the ROSAT PSPC detector. For the X-ray analysis we apply the method described in Uyan{\i}ker et al.~(1997). In the figures we only display the results for the best candidate; all results can be found in Table~\ref{tab1}.

\subsection{Spatial analysis}

\begin{figure}[htb] 
	\vspace{44mm} 
	\includegraphics{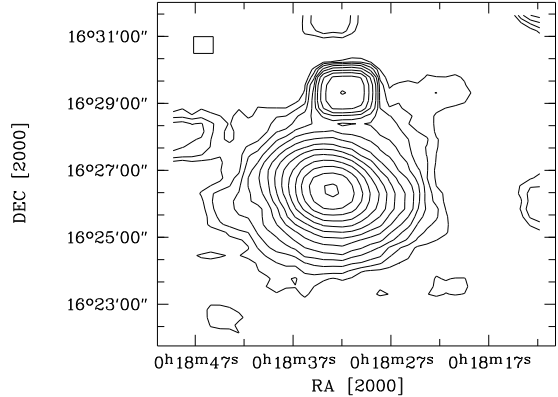}
\vspace{11mm}
\caption{ROSAT PSPC map of Cl0016+16. The spacing of the contour levels is 40\% of the value of the lower one; the value of the lowest line is $0.1\times 10^{-4} {\rm cts/sec/arcmin}^{2}$. The point source response has been removed before the spatial analysis. The rectangle in the upper left corner shows the resolution of the detector.\label{fig1}}
\end{figure}

The X-ray map of Cl0016+16 is shown in Fig.~\ref{fig1}. The contours are very symmetrically, the strong point source to the north is the quasar QSO~0015+162 (see Margon et al.~1983).

\begin{figure}[htb] 
	\vspace{42mm} 
	\includegraphics{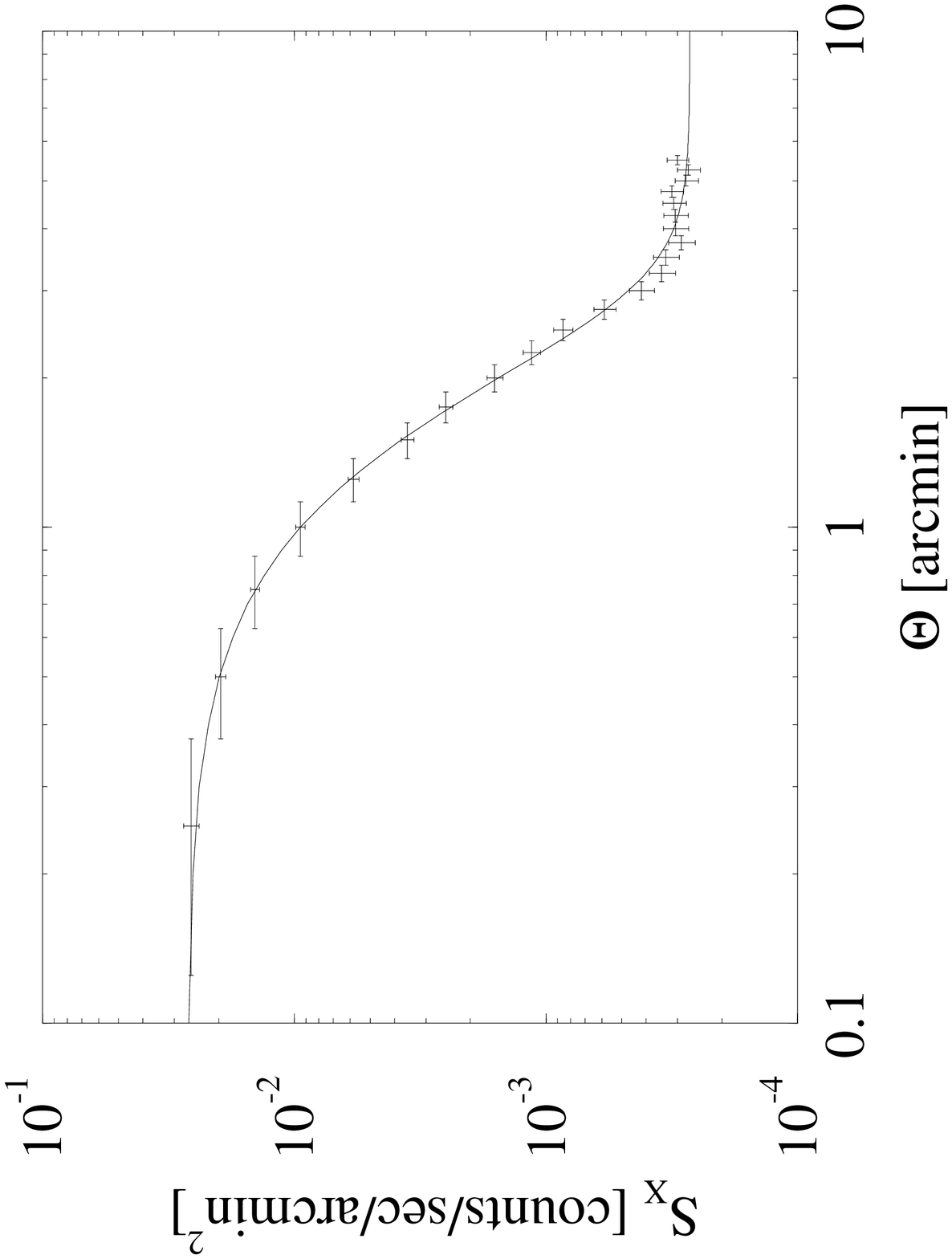}
\vspace{-40mm}

\hspace{45mm}{\footnotesize $
S_{\rm X}\!\left(\Theta\right) = 
S_{\rm X_{0}}\left[1+\left(\frac{\Theta}{\Theta_{\rm 
C}}\right)^{2}\right]^{-3\beta+\frac{1}{2}}+B
$}
\vspace{47mm}
\caption{Radial profile (data points) and the best fit (solid line) of the King model (equation) on the X-ray data for Cl0016+16. $B$ stands for a constant background.\label{fig2}}
\end{figure}

In the next step we plot an azimuthally averaged surface brightness profile, and obtain the spatial parameters of the electron distribution with the best fit for the King profile on the data. We determine the so-called core radius $\Theta_{\rm C}$, that describes the width of the distribution and the slope parameter $\beta$. In Fig.~\ref{fig2} the radial profile of Cl0016+16 with the best fit can be seen. For A85 we find evidence for the presence of a cooling flow. Here our model is insufficient: in a cooling flow cluster the temperature of the electrons is not constant. For future work on this cluster a different model is required.

\subsection{Spectral analysis}

For the analysis of spectral properties we use the theoretical model spectrum of the Raymond-Smith code. So we obtain the gas temperature $k_{{\rm B}}T_{{\rm e}}$ for the clusters (see Table~\ref{tab1}). Due to the moderate energy resolution of ROSAT the errors in the temperatures are large. Together with the small photon statistic for the Cl0016+16 observation we obtain a 90\% confidence error of about 37\%. As described in Uyan{\i}ker et al.~(1997) we calculate the luminosity distance $D_{\rm L}$, the linear extension of the core radius and the central electron density $n_{\rm e_{0}}$.

In Table~\ref{tab1} we summarize all obtained parameters with their 90\% confidence errors. We do not display the values for A85, since in that case the used model is inappropriate.

\begin{table}
\caption{Derived parameters from X-ray analysis with 90\% confidence 
errors. (with $H_{0} = 75$km/s/Mpc, $q_{0} = 0$)\label{tab1}}
\vspace*{1.5mm}
%\begin{center}
%\hspace*{-7mm}
\begin{tabular}{lcccccc}
\hline
\hline
& $\Theta_{\rm C}$ & $r_{\rm C}$ & $\beta$ & $kT_{\rm e}$ & $D_{\rm L}$ & $n_{\rm e_{0}}$ \\
& $[\,'\,]$ & $[{\rm kpc}]$ & & $[{\rm keV}]$ & $[{\rm Mpc}]$ & $[10^{-3}{\rm cm}^{-3}]$ \\
\hline
A665    & $1.43\pm 0.05$ & $236\pm 8\;\:$  & $0.64\pm 0.01$ & $6.3\pm 1.5$ & $791.8$ & $4.0\pm 0.5$ \\
Cl0016+16 & $1.63\pm 0.18$ & $550\pm 60$ & $1.26\pm 0.17$ & $7.3\pm 2.7$ & $2775$  & $5.6\pm 1.3$ \\
\hline
\end{tabular}
%\end{center}
\end{table}

\section{Calculation of the SZ diminution}

The change of the brightness temperature $T_{{\rm R}}$ towards clusters of galaxies at radio wavelength can be calculated as $\Delta T_{{\rm R}} = -2yT_{{\rm R}}$, where $y = \frac{\sigma_{{\rm T}}}{m_{{\rm e}} c^{2}} \int_{-\infty}^{\infty}{\rm d}l\,p_{{\rm e}}$ is the Comptonization parameter and $p_{\rm e} = n_{\rm e}k_{\rm B}T_{\rm e}$ the pressure of the intracluster electron cloud. With the isothermal $\beta$-model for the electron distribution the integration along the line of sight $l$ yields for the brightness temperature

\begin{equation}
\Delta T_{\rm R} = -Y_{0}T_{\rm R}\left[1+\left(\frac{\Theta}{\Theta_{\rm C}}\right)^{2}\right]^{-\frac{3 \beta}{2}+\frac{1}{2}}
\end{equation}
\noindent
with
\begin{equation}
Y_{0} = \frac{2\pi^{\frac{1}{2}}\sigma_{{\rm T}}k_{{\rm B}}T_{{\rm e}}n_{{\rm e}_{0}}r_{{\rm C}}}{m_{{\rm e}}c^{2}}\frac{\Gamma\left(\frac{3\beta}{2}-\frac{1}{2}\right)}{\Gamma\left(\frac{3\beta}{2}
\right)}.
\end{equation}

\begin{figure}[htb] 
	\vspace{45.5mm} 
	\includegraphics{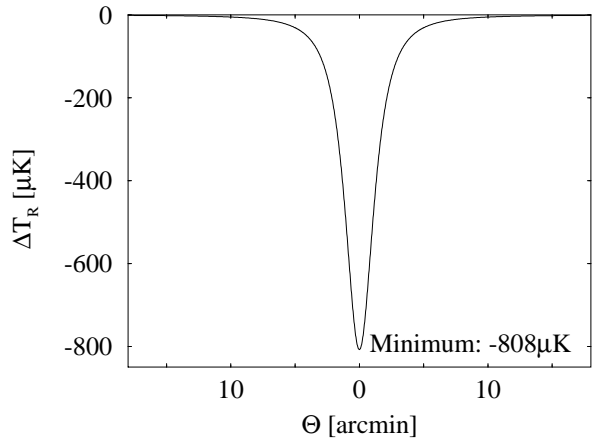}
\vspace{19mm}
\caption{Expected diminution of the temperature of the CMB towards Cl0016+16 with respect to the cluster center. $\Delta T_{\rm R} = 0$ means the undisturbed temperature of the CMB (2.73K). The calculated central diminution is in good agreement with recent observations (Carlstrom et al.~(1996))\label{fig3}}
\end{figure}

\noindent
To calculate the expected temperature distribution we use the parameters $T_{\rm e}$, $n_{\rm e_{0}}$, $r_{\rm C}$, $\Theta_{\rm C}$ and $\beta$ obtained from the X-ray analysis. For the integrated Comptonization parameter $Y_{0}$ we determine $1.6\pm 0.7$ for A665, and $3.0\pm 2.5$ for Cl0016+16, respectively. The central decrements of the brightness temperature $\Delta T_{\rm R,max}$ are derived as $-(440\pm 180)\mu {\rm K}$ for A665, and $-(810\pm 680)\mu {\rm K}$ for Cl0016+16. The 90\% confidence error ranges are large due to the moderate energy resolution of ROSAT, as mentioned above. In Fig.~\ref{fig3} the expected SZ diminution towards Cl0016+16 is shown. 

\section{Conclusions}

In this paper we have reported the X-ray analysis of the clusters of galaxies A85, A665, and Cl0016+16 in order to calculate the expected SZ diminution of the brightness temperature of the CMB towards these clusters. We assumed non-relativistic electrons and an isothermal King profile for the intracluster plasma distribution. The results show that measurements with the 100m telescope Effelsberg should be done only in case of Cl0016+16. From the X-ray analysis we found evidence for a cooling flow in A85.

Our group has proposed an observation of the SZ effect towards the cluster 0016+16 with the Effelsberg telescope. A more detailed description of this work is in preparation.

%-----------------------------------------------------------------
%  end of text
%-----------------------------------------------------------------
%-----------------------------------------------------------------
\end{document}